\documentclass[10pt, a4paper]{article}

\usepackage{lrec-coling2024} 
\usepackage{times}
\usepackage{latexsym}
\usepackage{graphicx}
\usepackage{booktabs}
\usepackage{multirow}
\usepackage{amsmath}
\usepackage{float}
\usepackage{pifont}
\usepackage{amssymb}
\usepackage{bbding}
\usepackage{soul}
\usepackage{algorithm}
\usepackage{algpseudocode}
\usepackage{adjustbox}
\usepackage[frozencache,cachedir=minted-cache]{minted}
\usepackage{subcaption}
\usepackage{graphicx}

\title{Structure-aware Fine-tuning for Code Pre-trained Models}

\name{Jiayi Wu\textsuperscript{1}\sthanks{~~Equal contribution}~, Renyu Zhu\textsuperscript{3}\footnotemark[1]~, Nuo Chen\textsuperscript{1}, Qiushi Sun\textsuperscript{4}, Xiang Li\textsuperscript{1}, Ming Gao\textsuperscript{1,2}\sthanks{~~Corresponding author}} 

\address{\textsuperscript{1}School of Data Science and Engineering, East China Normal University\\
\textsuperscript{2}KLATASDS-MOE in School of Statistics, East China Normal University\\
\textsuperscript{3}NetEase Fuxi AI Lab, \textsuperscript{4}National University of Singapore \\
         \{jiayiwu, nuochen\}@stu.ecnu.edu.cn, 
         zhurenyu@corp.netease.com\\
         qiushisun@u.nus.edu, \{xiangli,mgao\}@dase.ecnu.edu.cn}

\abstract{
Over the past few years, we have witnessed remarkable advancements in Code Pre-trained Models (CodePTMs). These models achieved excellent representation capabilities by designing structure-based pre-training tasks for code. However, how to enhance the absorption of structural knowledge when fine-tuning CodePTMs still remains a significant challenge. To fill this gap, in this paper, we present Structure-aware Fine-tuning (SAT), a novel structure-enhanced and plug-and-play fine-tuning method for CodePTMs.
We first propose a \textit{structure loss} to quantify the difference between the information learned by CodePTMs and the knowledge extracted from code structure.
Specifically, we use the attention scores extracted from Transformer layer as the learned structural information,
and the shortest path length between leaves in abstract syntax trees as the structural knowledge.
Subsequently, multi-task learning is introduced to improve the performance of fine-tuning.
Experiments conducted on four pre-trained models and two generation tasks demonstrate the effectiveness of our proposed method as a plug-and-play solution. 
Furthermore, we observed that SAT can benefit CodePTMs more with limited training data.
 \\ \newline  }

\begin{document}

\maketitleabstract

\section{Introduction}

Pre-trained language Models (PTMs)~\cite{liu2019roberta} with Transformer architecture \cite{vaswani2017attention} have greatly improved a wide range of natural language processing (NLP) tasks.
The success of PTMs in NLP also fosters the development of
Code Pre-trained Models (CodePTMs) for processing programming languages~\citep{sun2024ncisurvey}. 
Since \citet{feng2020codebert} proposed CodeBERT, a model with multi-layer bidirectional Transformer architecture pre-trained on NL-PL pairs, pre-training, and fine-tuning PTMs has gradually become the de facto paradigm for various code-related tasks leading to stunning performance.


Existing CodePTMs can be mainly categorized into two types: structure-free and structure-based models. 
The former, exemplified by CodeBERT~\cite{feng2020codebert}, regards code snippets as a sequence of tokens, thereby ignoring the inherent structure of code. The latter, such as GraphCodeBERT~\cite{guo2020graphcodebert} and CodeT5~\cite{wang2021codet5} both incorporate structure-based pre-training tasks to capture more structural knowledge.
Previous research~\cite{xu2022survey,chen2023pass,chen2023evaluating} has shown that structure-based CodePTMs that learn structural knowledge possess more substantial code representation capabilities. These structure-based CodePTMs mainly focus on adding structure-related tasks to learn general structural knowledge during pre-training. 
However, as demonstrated by \citet{wan2022they} and \citet{chen2022cat}, the CodePTMs tend to shift from general to task-specific structural knowledge in the fine-tuning stage. 
It further inspires us to think:
\textit{Can we enhance the absorption of task-specific structural knowledge in the fine-tuning stage for CodePTMs?}

\begin{figure}[t!]
    \centering
    \includegraphics[width=0.48\textwidth]{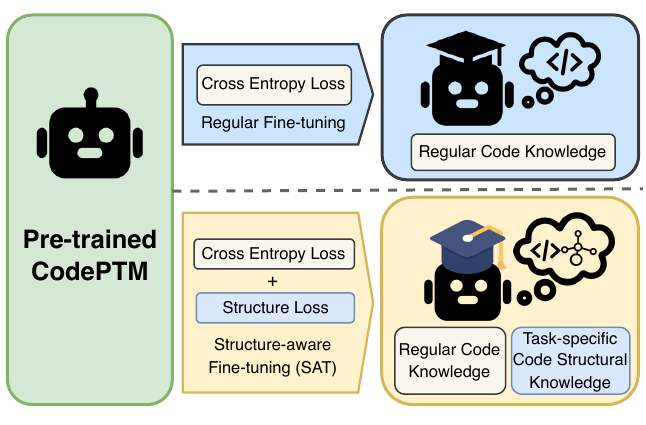}
    \caption{Compared to regular fine-tuning, CodePTMs can effectively capture task-specific code structure knowledge through structure-aware fine-tuning by structure loss.}
    \label{fig:SAT function}
\end{figure}

\begin{figure*}[t!]
     \centering
     \begin{subfigure}{0.33\textwidth}
        \centering
        \includegraphics[height=5.9cm]{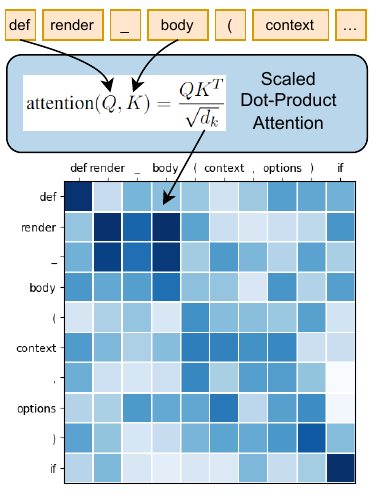}
        \caption{Generation of the attention matrix}
        \label{att}
     \end{subfigure}
     \hfill
     \begin{subfigure}{0.66\textwidth}
        \centering
        \includegraphics[height=5.7cm]{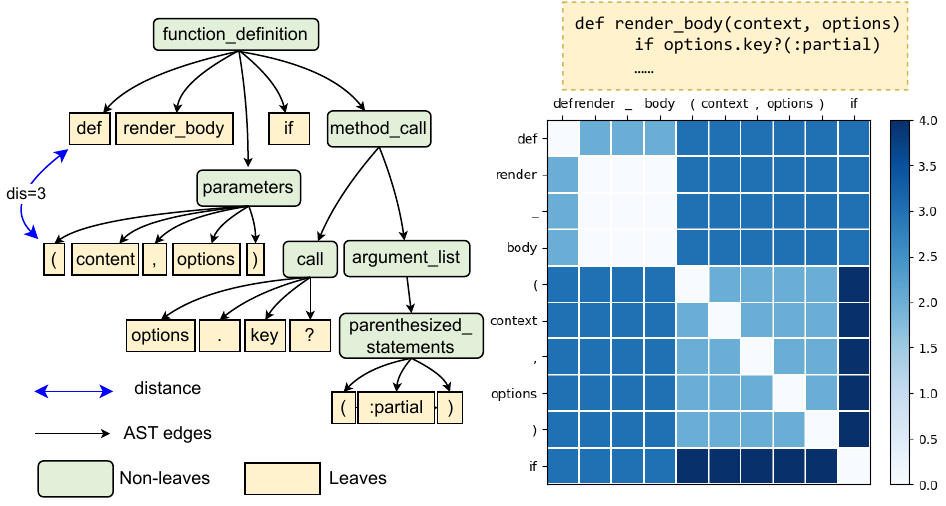}
        \caption{Generation of the distance matrix}
        \label{dis}
     \end{subfigure}
        \caption{Schematic illustration of generation of the attention matrix and the distance matrix. (a) Generation of the attention matrix. The code sequence is input into the multi-head self-attention block of CodePTMs, then the scaled dot-product values are computed to generate attention matrix. (b) Generation of the distance matrix. We first parse the raw code into AST, and calculate the shortest path between leaves on the AST to generate distance matrix. 
    }
        \label{fig:two mats}
\end{figure*}

Nevertheless, how to enhance the absorption of structural knowledge during the fine-tuning of CodePTMs still remains unaddressed~\citep{sun2024rethink}. Furthermore, most methods for learning structural code knowledge during the pre-training phase are not applicable during fine-tuning. For instance, if a fine-tuning approach were to imitate \citet{guo2022unixcoder} and \citet{Hybrid-DeepCom} by using traversed Abstract Syntax Trees (ASTs) as inputs, it would also necessitate constructing ASTs during inference.
Other methods by \citet{zugner2021language} and \citet{peng2021integrating} incorporate structural knowledge by modifying the attention mechanism within Transformer blocks, involve heavy model modifications and cannot be seamlessly applied to existing pre-trained models. Lastly, \citet{guo2020graphcodebert} and \citet{wang2021codet5} introduce structure-based pre-training tasks, which require predicting specific labels related to code structure, such as ``data flow edges'' or ``token identifier.'' However, these tasks are often unrelated to the fine-tuning tasks.

To address these issues, 
we present SAT, 
a \underline{s}tructure-\underline{a}ware fine-\underline{t}uning method for CodePTMs. As illustrated in Figure~\ref{fig:SAT function}, SAT can help CodePTMs to capture structural knowledge during the fine-tuning phase effectively and seamlessly. 
Firstly, we parse the code into an abstract syntax tree (AST) and construct a distance matrix by calculating the shortest path between two leaves in AST. 
Then, we extract the attention matrix from the Transformer layer of CodePTMs. 
After that, inspired by \citet{weng2019gan} and \citet{feydy2019interpolating}, 
we introduce the Sinkhorn divergence to compute the difference between the two matrices as a new objective function,
namely \textit{structure loss}. 
Finally, during the fine-tuning phase, we optimize the model by jointly optimizing the structure loss and the downstream task objective function using a multi-task approach.
We evaluate SAT on code summarization and translation tasks with four pre-trained models. Our contributions can be summarized as follows:

\begin{itemize}
    \item We propose SAT, a novel multi-task learning method that enhances the absorption of structural knowledge for CodePTMs during the fine-tuning stage. To the best of our knowledge, this is the first structure-aware fine-tuning method for CodePTMs.

    \item SAT can be easily implemented with the majority of Transformer-based CodePTMs as a plug-and-play solution.
    
    \item Our experiments involve four pre-trained models and two generation tasks, demonstrating that our proposed method can further improve the fine-tuning performance of CodePTMs. Additionally, SAT shows greater improvement in low-resource scenarios.
    
\end{itemize}

\section{SAT}

\begin{figure*}[t!]
    \centering
    \includegraphics[width=0.83\textwidth]{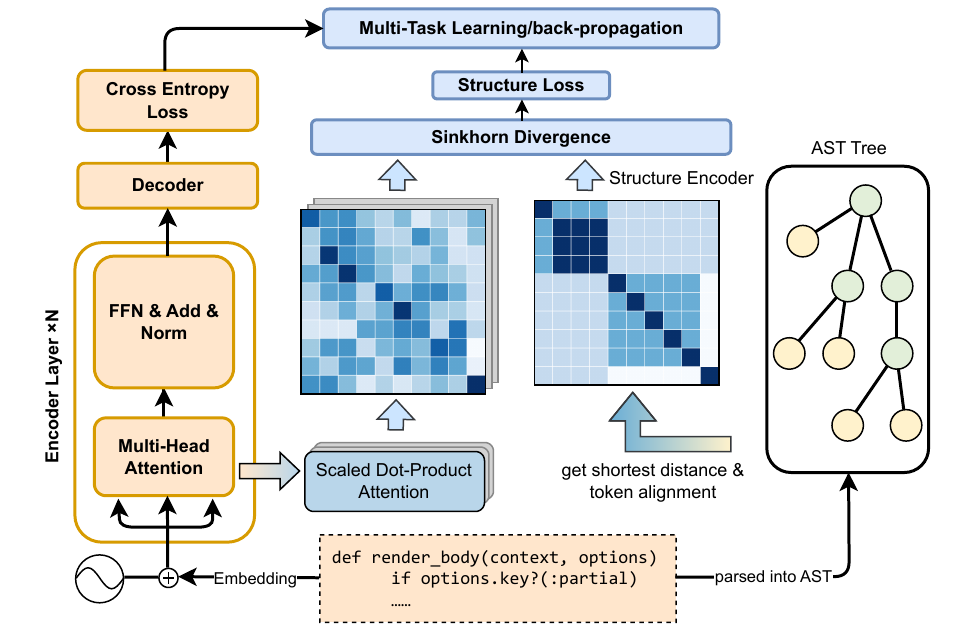}
    \caption{The architecture of the SAT method. Given a code snippet, we parsed it into an AST, and the distance matrix is obtained by computing the shortest distances between leaf nodes. The structural information is extracted using the struct encoder. The code snippet is then input into the CodePTMs, from which the attention matrix is extracted from the multi-head attention blocks. The sinkhorn divergence between the distance matrix and the attention matrix is calculated to obtain the structure loss.}
    \label{fig:SAT overall}
\end{figure*}

In this section, we introduce SAT.
The detailed architecture of the SAT is illustrated in Figure~\ref{fig:SAT overall}.

\subsection{Code Basics}
As shown in Figure~\ref{dis}, AST is a tree-like model used to represent the structure of a program, abstracting the source code into nodes and connecting edges. In this paper, we use Tree-sitter~\footnote{\href{https://github.com/tree-sitter}{github.com/tree-sitter}} to generate ASTs.

\subsection{Distance Matrix and Attention Matrix}
\citet{chen2022cat} proposed that the model’s performance improves with its ability to explore structural knowledge, leading to a greater alignment between CodePTMs’ token-level attention scores and AST nodes’ pair-wise distances. Inspired by this work, we aim to quantify the disparity between the learned structural information by the pre-trained models and the structural knowledge from AST. To achieve this, we introduce the attention matrix and the distance matrix, respectively.

Figure~\ref{fig:two mats} illustrates the process of generating the attention matrix and the distance matrix. The generation of the attention matrix is shown in Figure~\ref{att}, the raw code is input into the multi-head self-attention block of CodePTMs as a sequence of subtokens, where the scaled dot-product attention between subtokens is calculated. These attention values are extracted and used to construct the attention matrix.
And Figure~\ref{dis} illustrates the generation of the distance matrix. 
We parse the raw code into an AST, where each leaf node is derived from the raw code. To calculate the distances between tokens, we treat the AST as an undirected graph, where the distance between tokens corresponds to the length of the shortest path between the corresponding nodes in the AST. For example, the distance between token ``def'' and token ``('' is $3$. 
We extract the shortest distances between tokens from the AST to construct the distance matrix.

So far, the shapes of the attention matrix and the distance matrix are different. The attention matrix is at the subtoken level, while the distance matrix is at the token level. Subtokens are determined by the tokenizer, while tokens are determined by the tree-sitter. To facilitate the comparison between the distance matrix and the attention matrix, we convert the distance matrix from the token level to the subtoken level, ensuring that both matrices have the same shape. Specifically, subtokens are generally contained within tokens, as shown in Figure~\ref{dis}, subtoken ``render'', ``\_'' and ``body'' correspond to token ``render\_body''. By iterating through the sequences of subtokens and tokens, we get the mapping relationship between subtoken and token. The distance between subtokens is the distance between the corresponding tokens. In Figure~\ref{dis}, for example, the subtoken ``render'' is contained within the token ``render\_body'', and the subtoken ``:'' is contained within the token ``:partial''. The distance between them is $5$.

\subsection{Formal Definition of Structure Loss}
\label{def strcture loss}

Inspired by \citet{weng2019gan} and \citet{feydy2019interpolating}, we employ Wasserstein distance to quantify the differences between the attention matrix and distance matrix. The Wasserstein distance is a smooth metric for measuring the distance between two probability distributions, even when these distributions reside in non-overlapping lower dimensional manifolds \cite{weng2019gan}. However, computing the Wasserstein distance entails high computational costs. To mitigate this, we utilize Sinkhorn divergences, which are positive and definite approximations of Wasserstein distances, and offer efficient batch computation on GPUs \cite{feydy2019interpolating}. In this paper, we leverage the Geomloss~\footnote{\href{https://www.kernel-operations.io/geomloss/index.html}{https://www.kernel-operations.io/geomloss/index.html}} for calculating Sinkhorn divergences.

Because the values of scaled dot-product attention and distance are not on the same scale, Sinkhorn divergences are not suitable for measuring the differences between two matrices that have values in different orders of magnitude. 
To address this issue, we introduce a linear layer that applies coefficients and biases to the distance matrix~\cite{liu2012robust}. This linear layer, referred to as the structure encoder, is treated as learnable parameters integrated into the model.

We treat the attention matrix and the distance matrix as two distributions. By applying the structure encoder to the distance matrix for scaling and biasing, we utilize Sinkhorn divergences to measure the differences between the two matrices to obtain the \textit{structure loss}:
\begin{equation}
    \mathcal{L}_{i}=\text{Sinkhorn divergences}(\mathbf{A},\text{Linear}(\mathbf{D})),
    \label{eq:structure loss}
\end{equation}
\begin{equation}
    \mathcal{L}_{\text{structure}}=\frac{\sum_{i=1}^{h}{\mathcal{L}_{i}}}{h}
\end{equation}
where $\mathbf{A},\mathbf{D}\in \mathbf{\mathbb{R}}^{n \times n}$, $\mathbf{A}$ denotes the attention matrix,  $\mathbf{D}$ represents the distance matrix, $n$  represents the length of the input sequence in CodePTMs, and $h$ represents the number of attention heads.

Among the Transformer layers in CodePTMs, the first layer captures the most structural information~\cite{chen2022cat}. Therefore, we extract the attention values from the first Transformer layer and utilize them as the attention matrix.

We draw inspiration from the concept of multi-task fine-tuning \cite{kalyan2021ammus} to jointly optimize the objective function of downstream tasks and the structure loss. This approach enables the model to learn the structural information of the code during the backpropagation phase. The final objective function is followed as: 
\begin{equation}
    \mathcal{L}_{\text{final}}=\mathcal{L}_{\text{task}}+\alpha \cdot \mathcal{L}_{\text{structure}}
    \label{eq:multi-task}
\end{equation}
where $\mathcal{L}_{task}$ represents the objective function of the downstream task, such as cross entropy loss for code translation and code summarization. And $\alpha$ is a hyperparameter that controls the weight of the structure loss.

As a plug-and-play method, the \textit{structure loss} can be easily applied when needed and is applicable to majority Transformer-based CodePTMs, demonstrating its generality.

\section{Experiments}

\subsection{Tasks and Metrics}

To evaluate the ability of our proposed method SAT, we select two challenging tasks in the CodeXGLUE \cite{CodeXGLUE} Benchmark: code summarization and code translation, to assess cross-modal generation and code-to-code generation capabilities, respectively.

Code summarization aims to generate a natural language description for a given programming language snippet. We evaluated the effectiveness of our method using the CodeSearchNet \cite{CodeSearchNet} dataset for code summarization across five programming languages: Ruby, JavaScript, Go, Python, and Java. We evaluate the results using the smoothed BLEU-4~\cite{lin2004orange} metric.

Code translation involves translating a code from one programming language to a different one. We conducted experiments on this task using the dataset provided by CodeXGLUE, which consists of ten thousand pairs of Java and C\# code snippets with equivalent functionality. We follow the setting of CodeXGLUE and employ BLEU-4 and exact match (EM) as evaluation metrics.

\subsection{Backbone models}

We implement SAT to four representative PTMs, encompassing two distinct types: encoder-only and encoder-decoder models. 

For the encoder-only architecture, we employed RoBERTa~\cite{liu2019roberta}, a pre-trained model trained on a large-scale text corpus using masked language modeling (MLM). 
Additionally, we choose two RoBERTa-based CodePTMs, CodeBERT~\cite{feng2020codebert} and GraphCodeBERT~\cite{guo2020graphcodebert}.

In the case of the encoder-decoder architecture, we employ CodeT5~\cite{wang2021codet5}, a state-of-the-art CodePTM based on the T5 framework~\cite{raffel2020exploring}, known for its exceptional performance on various tasks within the CodeXGLUE benchmark~\citep{CodeXGLUE}. 

\subsection{SAT Effectiveness}
\label{sec:3.3}

\begin{table*}
\centering
\resizebox{\textwidth}{!}{
\small
    \begin{tabular}{l | c c c c c c | c c c c}
    \toprule
    \multirow{3}{*}{\textbf{Model}} & \multicolumn{6}{c}{\textbf{Code summarization (Smoothed BLEU-4)}} & \multicolumn{2}{c}{\textbf{Java to C\#}} & \multicolumn{2}{c}{\textbf{C\# to Java}}\\
    \cmidrule(lr){2-7} \cmidrule(lr){8-9} \cmidrule(lr){10-11}
    & \textbf{Ruby} & \textbf{JavaScript} & \textbf{Go} & \textbf{Python} & \textbf{Java} & \textbf{Overall} & \textbf{BLEU} & \textbf{EM}  & \textbf{BLEU} & \textbf{EM}\\
    \midrule
    RoBERTa & 11.17 & 11.90 & 17.72 & 18.14 & 16.47 & 15.08 & 77.46 & 56.10  & 71.99 & 57.90\\
    RoBERTa+SAT & \textbf{11.82} & \textbf{15.68} & \textbf{17.85} & \textbf{18.49} & \textbf{18.21} & \textbf{16.41} & \textbf{80.28} & \textbf{59.30} & \textbf{76.67} & \textbf{61.00}\\
     & (+0.65) & (+3.78) & (+0.13) & (+0.35) & (+1.74) & (+1.33) & (+2.82) & (+3.20) & (+4.68) & (+3.10) \\
    \midrule
    CodeBERT & 12.16 & 14.90 & 18.07 & \textbf{19.06} & 17.65 & 16.37 & 79.92 & 59.00 & 72.14 & 58.80\\
    CodeBERT+SAT & \textbf{12.50} & \textbf{16.34} & \textbf{18.32} & 19.02 & \textbf{18.64} & \textbf{16.96} & \textbf{80.51} & \textbf{61.30} & \textbf{76.40} & \textbf{62.00}\\
     & (+0.34) & (+1.44) & (+0.25) & (-0.04) & (+0.99) & (+0.59) & (+0.59) & (+2.30) & (+4.26) & (+3.20)\\
    \midrule
    GraphCodeBERT & 12.39 & 14.81 & 18.41 & 18.06 & 19.00 & 16.53 & 80.58 & 59.40 & 72.64 & 58.80\\
    GraphCodeBERT+SAT & \textbf{13.16} & \textbf{16.33} & \textbf{18.63} & \textbf{19.24} & \textbf{19.15} & \textbf{17.30} & \textbf{81.74} & \textbf{62.30} & \textbf{77.30} & \textbf{61.50}\\
     & (+0.77) & (+1.52) & (+0.22) & (+1.18) & (+0.15) & (+0.77) & (+1.16) & (+2.90) & (+4.66) & (+2.70)\\
    \midrule
    CodeT5 & 15.24 & 16.16 & 19.56 & 20.01 & 20.31 & 18.25 & 84.03 & 65.90 & 79.87 & 66.90\\
    CodeT5+SAT & \textbf{15.51} & \textbf{16.20} & \textbf{19.73} & \textbf{20.34} & \textbf{20.48} & \textbf{18.45} & \textbf{84.99} & \textbf{67.50} & \textbf{80.33} & \textbf{67.70}\\
     & (+0.27) & (+0.04) & (+0.17) & (+0.33) & (+0.17) & (+0.20) & (+0.96) & (+1.60) & (+0.46) & (+0.80)\\
    \bottomrule
    \end{tabular}
}
    \caption{\label{result}
    The results of SAT on code summarization and code translation tasks. The left half shows the smoothed BLEU-4 scores on the code summarization tasks, while the right half shows the BLEU-4 scores and exact match (EM) accuracy for code translation tasks. ``+SAT'' indicates that the model was enhanced with the SAT method. Better results in the same task are bolded.
    }
\end{table*}

\begin{figure*}[t!]
     \centering
     \begin{subfigure}{0.32\textwidth}
        \centering
        \includegraphics[height=5cm]{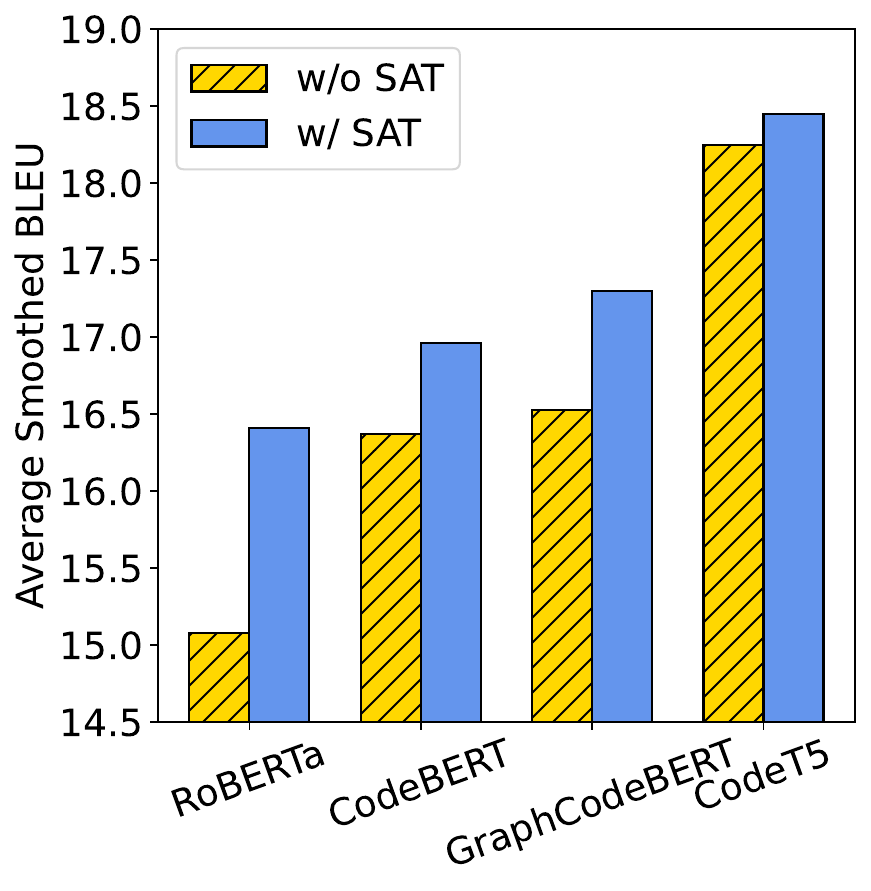}
        \caption{The average smoothed BLEU-4 scores on the code summarization}
     \end{subfigure}
     \hfill
     \begin{subfigure}{0.32\textwidth}
        \centering
        \includegraphics[height=5cm]{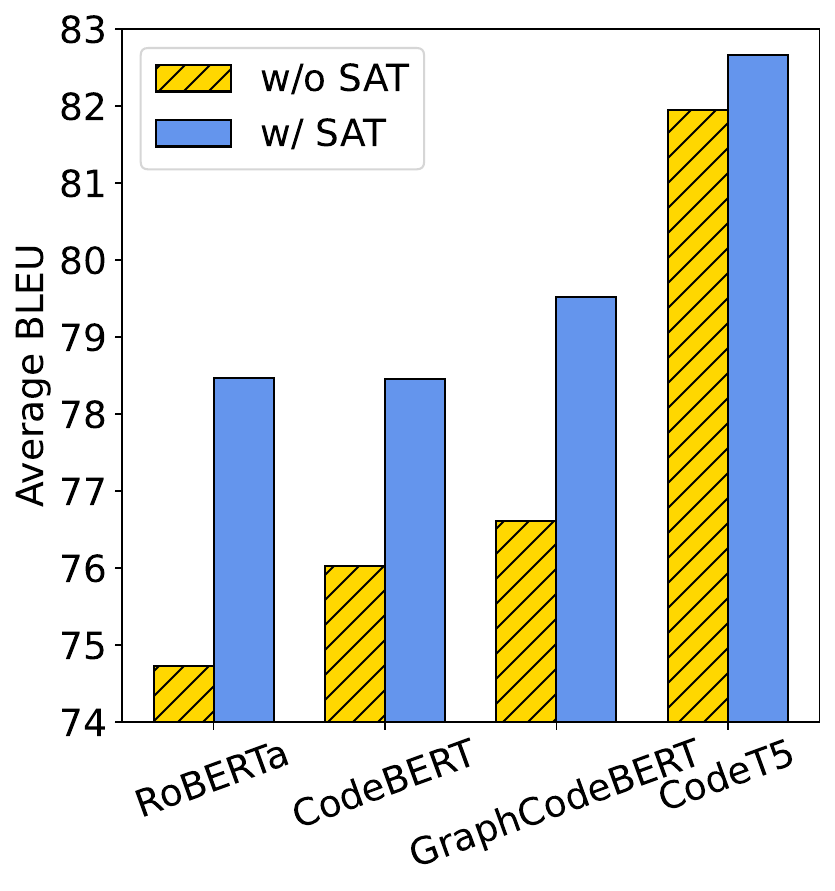}
        \caption{The average BLEU-4 scores for the code translation}
     \end{subfigure}
     \hfill
     \begin{subfigure}{0.32\textwidth}
        \centering
        \includegraphics[height=5cm]{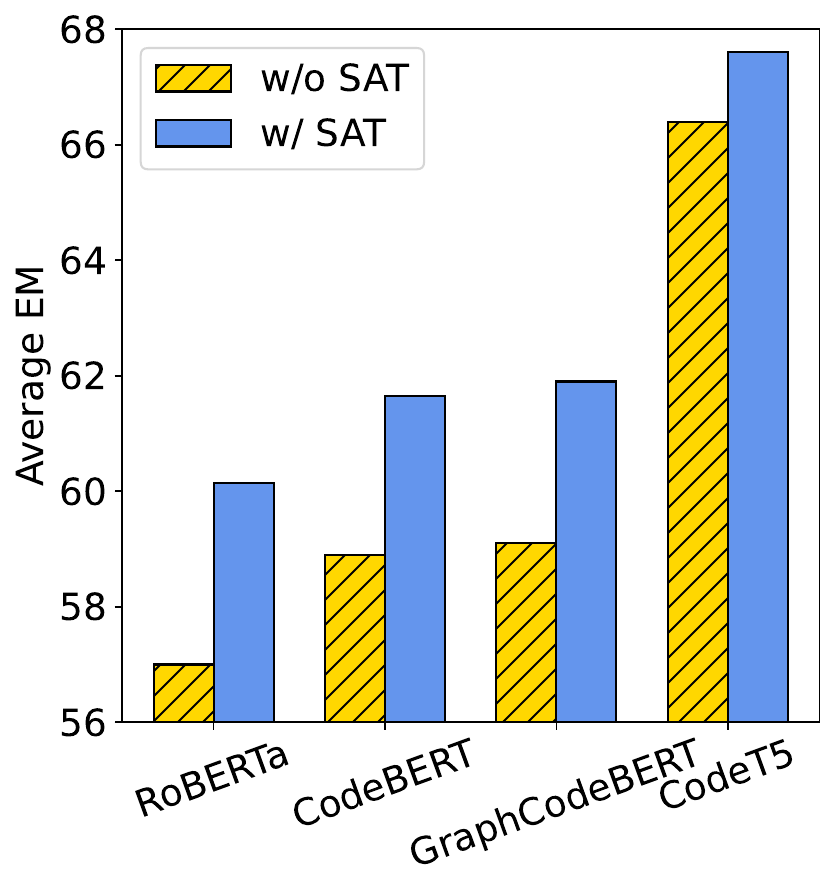}
        \caption{The average exact match (EM) for the code translation.}
     \end{subfigure}
        \caption{Average results of SAT on code summarization and code translation tasks.}
    \label{fig:line sat}
\end{figure*}

The results of the SAT method on the code summarization and code translation tasks are presented in Table~\ref{result}. Upon analyzing these experimental results, we discover that: 
\begin{itemize}
    \item \textbf{SAT can enhance the absorption of structural knowledge for CodePTMs in the fine-tuning stage}:
    Our proposed SAT approach obtains consistent improvements across the RoBERTa, CodeBERT, GraphCodeBERT, and CodeT5 models. In some scenarios, weaker models enhanced by our method can outperform stronger models, e.g., on code summarization task, RoBERTa enhanced by SAT even surpasses the unaugmented CodeBERT. Similarly, CodeBERT's performance after SAT augmentation outperforms unaugmented GraphCodeBERT. Notably, significant performance gains are observed in some models and tasks. For instance, RoBERTa+SAT achieves a smooth BLEU-4 score improvement of $3.78$ on the JavaScript dataset for code summarization. And for the task of translating C\# to Java, RoBERTa+SAT achieves an exact match improvement of $4.68$.

    \item \textbf{SAT can improve more with weaker models}:
    Figure~\ref{fig:line sat} illustrates the performance improvement of the SAT method across different backbone models. Among these models, RoBERTa exhibits the most substantial enhancement, followed by CodeBERT and GraphCodeBERT, while CodeT5 shows the least improvement. 
    This is likely due to the fact that RoBERTa is trained on pure textual corpora and lacks explicit code structural knowledge during the pre-training phase. Consequently, incorporating SAT during fine-tuning significantly enhances the model's performance. In contrast, CodeT5 already acquires extensive structural code information during pre-training through identifiers modeling, resulting in a smaller impact from SAT. These findings demonstrate that SAT effectively assists PTMs in learning structural code knowledge during fine-tuning, with models possessing less pre-trained code knowledge experiencing more notable performance gains.

\end{itemize}

\subsection{Exploration of Different Training Data Scales}
\begin{table*}[htb]
\centering
\resizebox{\textwidth}{!}{
    \begin{tabular}{c|l c c c c|l c c c c}
    \toprule
    \multirow{2}{*}{\textbf{Sample rate}} & \multirow{2}{*}{\textbf{Model}} & \multicolumn{2}{c}{\textbf{Code summarization}} & \multicolumn{2}{c}{\textbf{Code Translation}} & \multirow{2}{*}{\textbf{Model}} & \multicolumn{2}{c}{\textbf{Code summarization}} & \multicolumn{2}{c}{\textbf{Code Translation}} \\
    \cmidrule(r){3-4} \cmidrule(r){5-6} \cmidrule(r){8-9} \cmidrule(r){10-11} 
    & & \textbf{Ruby} & \textbf{Python} & \textbf{BLEU} & \textbf{EM} & & \textbf{Ruby} & \textbf{Python} & \textbf{BLEU} & \textbf{EM} \\
    \midrule 
    \multirow{3}{*}{20\%} 
     & CodeBERT &  11.64 & 18.66 & 75.66 & 56.10 & CodeT5 &  15.18 & 19.90 & 81.21 & 61.10 \\
     & CodeBERT+SAT &  11.98 & 18.98 & 76.58 & 57.20 & CodeT5+SAT &  15.89 & 20.27 & 81.97 & 61.90 \\
     & & \textbf{(+0.34)} & \textbf{(+0.32)} & \textbf{(+0.92)} & (+1.10) & & \textbf{(+0.71)} & \textbf{(+0.37)} & \textbf{(+0.76)} & \textbf{(+0.80)} \\
    \midrule
    \multirow{3}{*}{40\%} 
     & CodeBERT &  12.09 & 18.98 & 78.76 & 59.00 & CodeT5 &  15.24 & 19.96 & 84.13 & 65.80 \\
     & CodeBERT+SAT & 12.27 & 19.00  & 79.55 & 60.40 & CodeT5+SAT & 15.66 & 20.17 & 84.52 & 66.00 \\
     & & (+0.18) & (+0.02) & (+0.79) & \textbf{(+1.40)} & & (+0.42) & (+0.21) & (+0.39) & (+0.2) \\
    \midrule
    \multirow{3}{*}{60\%} 
     & CodeBERT &  12.16 & 18.87 & 79.39 & 59.40 & CodeT5 & 15.19 & 20.18 & 84.41 & 66.20 \\
     & CodeBERT+SAT &  12.37 & 19.03 & 80.03 & 59.50 & CodeT5+SAT & 15.35 & 20.24 & 84.58 & 66.80 \\
     & & (+0.21) & (+0.16) & (+0.64) & (+0.10) & & (+0.16) & (+0.06) & (+0.17) & (+0.60) \\
    \midrule
    \multirow{3}{*}{80\%} 
     & CodeBERT & 11.87 & 18.99 & 80.32 & 61.50 & CodeT5 & 15.36 & 20.20 & 84.68 & 66.40 \\
     & CodeBERT+SAT & 11.75 & 19.11 & 80.88 & 61.90 & CodeT5+SAT & 15.56 & 20.21 & 84.77 & 66.20 \\
     & & (-0.12) & (+0.12) & (+0.56) & (+0.40) & & (+0.20) & (+0.01) & (+0.09) & (-0.20) \\
    \bottomrule
    \end{tabular}
    } 
    \caption{\label{sample}
    Result of fine-tuning on a limited training dataset. The highest performance improvement is bolded. (Code Translation refers to the task of translating Java to C\#.)
    }
\end{table*}
\begin{figure*}[htp]
     \centering
     \begin{subfigure}{0.32\textwidth}
        \centering
        \includegraphics[height=4.2cm]{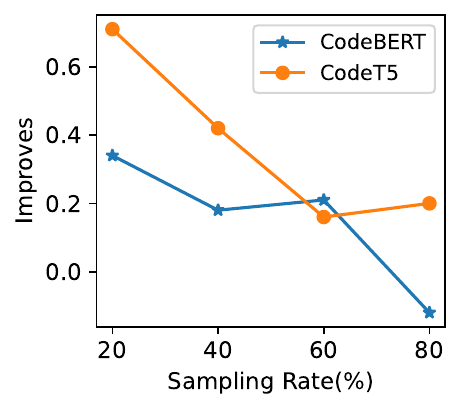}
        \caption{Summarization on Ruby}
     \end{subfigure}
     \hfill
     \begin{subfigure}{0.32\textwidth}
        \centering
        \includegraphics[height=4.2cm]{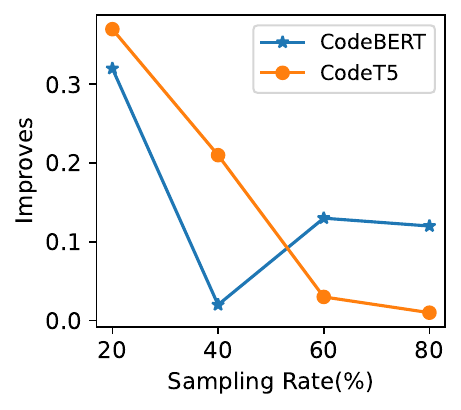}
        \caption{Summarization on Python}
     \end{subfigure}
     \hfill
     \begin{subfigure}{0.32\textwidth}
        \centering
        \includegraphics[height=4.2cm]{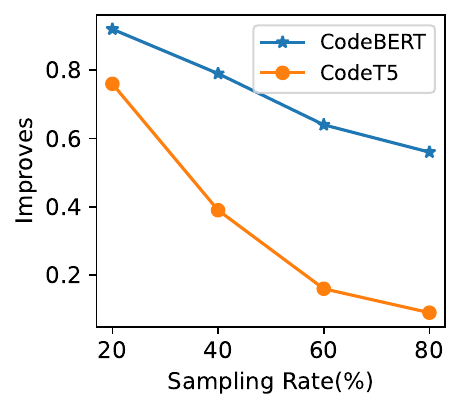}
        \caption{Translate Java to C\#}
     \end{subfigure}
        \caption{Performance improvement of fine-tuning with SAT on limited training datasets.}
        \label{fig:sample}
\end{figure*}
In order to evaluate the ability of the SAT method to extract code structural information from limited data, we conducted experiments using both the CodeBERT and CodeT5 models on code summarization and code translation tasks with different training data scale. We sampled 20\%, 40\%, 60\%, and 80\% of the training data, and the results are presented in Table 2. 
From the results, we observe that the SAT method consistently enhances the performance of the model across various training data scales.

Significantly, as shown in Figure~\ref{fig:sample}, we observe that \textbf{the SAT method yields more substantial performance improvements as the training data scale decreases}. We believe this is because, under the influence of the SAT method, the language model can learn a sufficient amount of structural information without requiring a large amount of training data.
This ability becomes evident when the language model is fine-tuned with a small-scale dataset, as it can learn a considerable amount of code structural information even in the absence of sufficient semantic knowledge, thus compensating for the limitations in semantic understanding. This indicates that SAT can better assist models in enhancing performance in low-resource scenarios.

\subsection{Did the SAT really learn the structural information?}

\paragraph{Analysis from the perspective of structure loss}

\begin{figure*}[t]
     \centering
     \begin{subfigure}{0.32\textwidth}
        \centering
        \includegraphics[height=4.2cm]{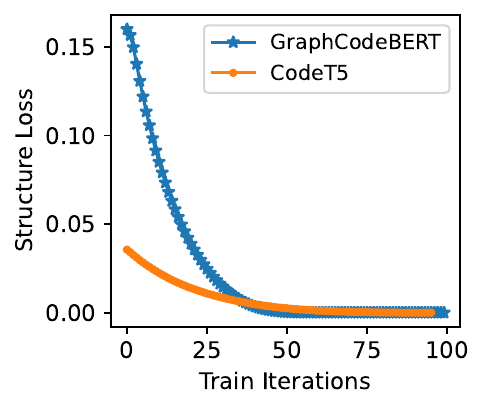}
        \caption{Translate C\# to Java}
     \end{subfigure}
     \hfill
     \begin{subfigure}{0.32\textwidth}
        \centering
        \includegraphics[height=4.2cm]{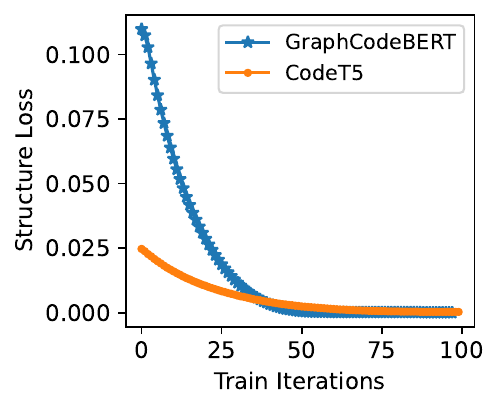}
        \caption{Translate Java to C\#}
     \end{subfigure}
     \hfill
     \begin{subfigure}{0.32\textwidth}
        \centering
        \includegraphics[height=4.2cm]{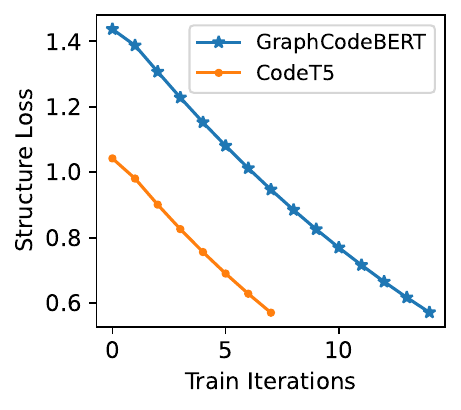}
        \caption{Summarizaiton on Ruby}
     \end{subfigure}
        \caption{Curve of structure loss}
        \label{fig:structure loss}
\end{figure*}

We optimize the structure loss by multi-task fine-tuning. The curve of structure loss with the number of training iterations is shown in Figure~\ref{fig:structure loss}. We observe that the structure loss gradually decreases as the number of iterations increases. This indicates a diminishing difference between the attention and distance matrices, suggesting that the model is learning structural information. Notably, in the early stages of fine-tuning, the structure loss of CodeT5+SAT is consistently lower than that of GraphCodeBERT+SAT. This observation indicates that CodeT5 has learned more structural knowledge during the pre-training phase.

\begin{figure}[t!]
     \centering
     \begin{subfigure}{0.49\textwidth}
        \centering
        \includegraphics[height=4cm]{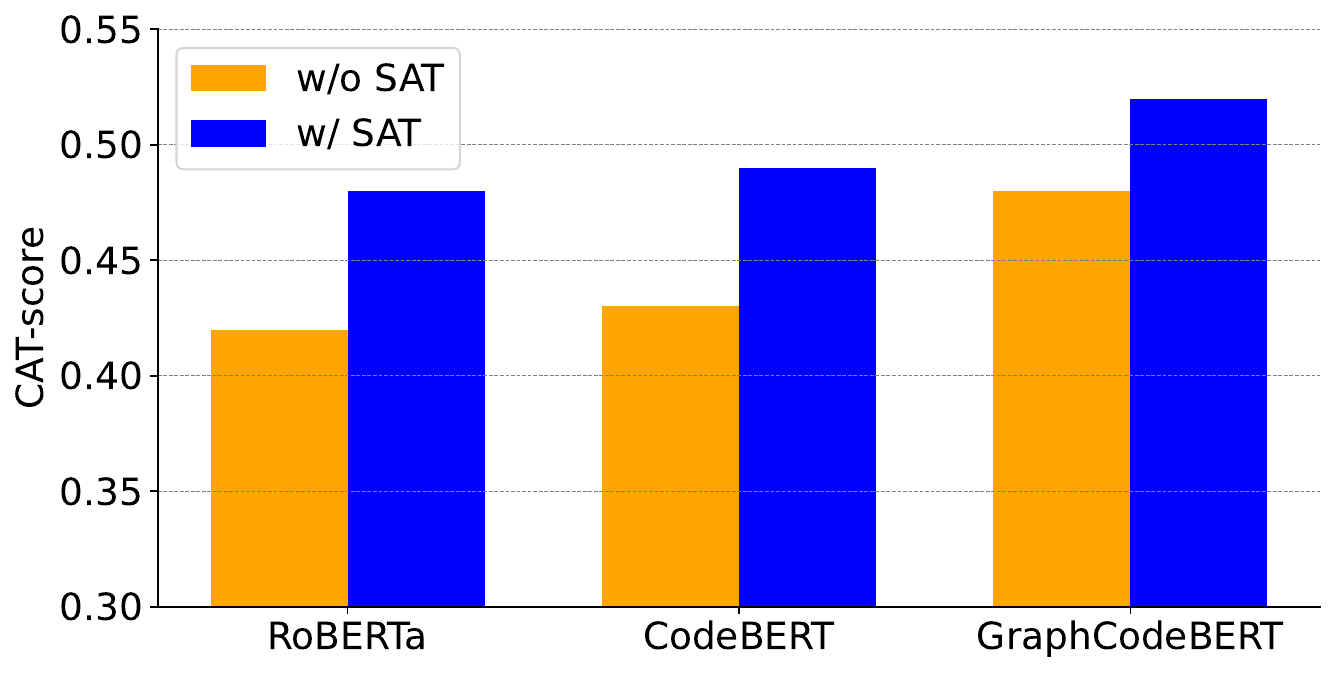}
        \caption{Summarization on Ruby}
     \end{subfigure}
     \begin{subfigure}{0.49\textwidth}
        \centering
        \includegraphics[height=4cm]{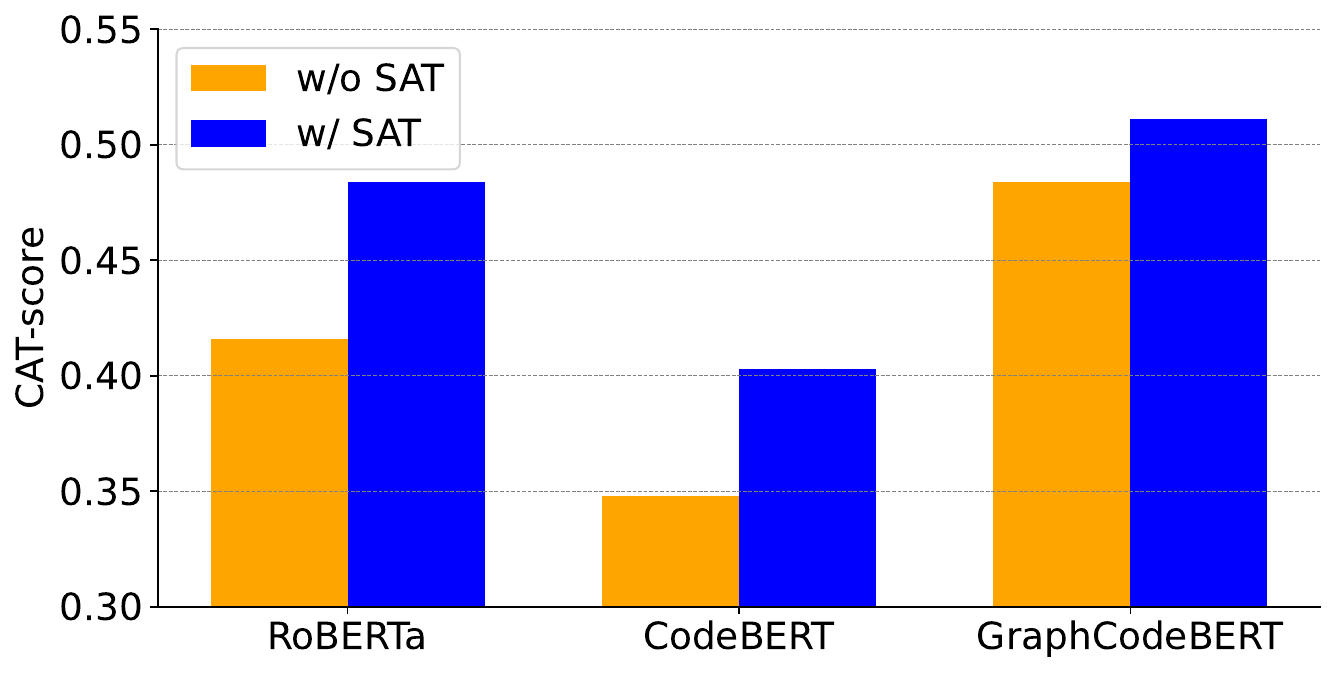}
        \caption{Summarization on Java}
     \end{subfigure}
     \caption{Comparative of CAT Scores with and without the SAT}
        \label{fig:catscore}
\end{figure}

\paragraph{Analyzing from CAT-probing}
CAT-probing \cite{chen2022cat} is a probing method to quantitatively interpret how CodePTMs attend code structure. It defines a new metric CAT-score to measure the commonality between the token-level attention scores generated in CodePTMs and the pair-wise distances between corresponding AST nodes. The higher the CAT-score, the stronger the ability of CodePTMs to capture code structure.

To assess whether SAT enhances the absorbtion of code structural knowledge, we selected the RoBERTa, CodeBERT, and GraphCodeBERT models and conducted code summarization tasks on Ruby and Java datasets and computed the CAT-score for each model with and without the incorporation of SAT.

As shown in Figure~\ref{fig:catscore}, we observed that all models exhibited an increase in their CAT-scores upon the introduction of SAT. This indicates that SAT has the ability to enhance pre-trained models to capture a greater amount of code structural knowledge. The degree of improvement in CAT-scores is correlated with the level of code knowledge acquired during the pre-training phase by the backbone models, with RoBERTa exhibiting the greatest improvement, followed by CodeBERT and GraphCodeBERT. This indicates that SAT is particularly effective in assisting models with relatively weaker performance, thus compensating for their deficiency in acquiring an adequate level of code structural knowledge during the training phase. This observation aligns with the findings presented in Section \ref{sec:3.3}.

\paragraph{Analyzing from a case study}\
\begin{table*}
  \centering
    \small
  \begin{tabular}{c|c|c}
    \toprule
     & \textbf{Case 1} & \textbf{Case 2} \\
    \midrule
    Source &
    \begin{minipage}{0.3\textwidth}
\begin{minted}[escapeinside=||,fontsize=\tiny]{csharp}
public override V next() 
{
    return this.nextEntry().value;
}
  \end{minted}
\end{minipage}
    &
    \begin{minipage}{0.5\textwidth}
\begin{minted}[escapeinside=||,fontsize=\tiny]{csharp}
public SinglePositionTokenStream(string word)
{
    termAtt = addAttribute(CharTermAttribute.class);
    posIncrAtt = addAttribute(PositionIncrementAttribute.class);
    this.word = word;
   
  \end{minted}
\end{minipage}
    \\
    \midrule
    Target &
    \begin{minipage}{0.3\textwidth}
\begin{minted}[escapeinside=||,fontsize=\tiny]{csharp}
public V next()
{
    return super.nextEntry().getValue();
}
  \end{minted} 
\end{minipage}
    & 
    \begin{minipage}{0.5\textwidth}
\begin{minted}[escapeinside=||,fontsize=\tiny]{java}
public SinglePositionTokenStream(String word) 
{
    termAtt = addAttribute(CharTermAttribute.class);
    posIncrAtt = addAttribute(PositionIncrementAttribute.class);
    this.word = word;
    returned = true;
}
  \end{minted} 
\end{minipage}
    \\
    
    \midrule
    RoBERTa &
    \begin{minipage}{0.3\textwidth}
\begin{minted}[escapeinside=||,fontsize=\tiny]{csharp}
public V next()
{
    V v = nextValue;
    advance();
    return v;
}
  \end{minted} 
\end{minipage}
    & 
    \begin{minipage}{0.5\textwidth}
\begin{minted}[escapeinside=||,fontsize=\tiny]{java}
public SinglePositionTokenStream(String word) 
{
    termAtt = addAttribute(CharTermAttribute.class);
    posIncrAtt = word;
    returned = true;
}
  \end{minted} 
\end{minipage}\\

\midrule
    \textbf{\shortstack{RoBERTa\\+SAT}}  &
    \begin{minipage}{0.3\textwidth}
\begin{minted}[escapeinside=||,fontsize=\tiny]{csharp}
public V next() 
{ 
    return super.nextEntry().value; 
}
  \end{minted} 
\end{minipage}
    & 
    \begin{minipage}{0.5\textwidth}
\begin{minted}[escapeinside=||,fontsize=\tiny]{java}
public SinglePositionTokenStream(String word) 
{
    termAtt = addAttribute(CharTermAttribute.class);
    posIncrAtt = addAttribute(PositionIncrementAttribute.class);
    this.word = word;
    returned = true;
}
  \end{minted} 
\end{minipage}
    \\

    \bottomrule
  \end{tabular}
  \caption{\label{tab:case study}
  Case study}
\end{table*}
We select two representative cases to illustrate the difference in model outputs with and without the SAT method. As shown in Table~\ref{tab:case study}, we provide two examples of the test data of translating C\# to Java. In the first example, the output structure with the SAT method is consistent with the correct answer, except for the token ``value'', which differs from the correct answer, while the output structure without the SAT method is entirely dissimilar to the correct answer. In the second example, the output with the SAT method is identical to the correct answer, whereas the output without the SAT method has misaligned variable assignments, indicating a lack of learned structural information from the input code sequence. These two examples demonstrate that SAT aids CodePTMs in acquiring more structural knowledge.
\section{Related Work}
\subsection{Pre-Trained Language Models for Programming Languages}
\label{rela:codeptms}
With the great success of pre-trained language models in natural language processing, transformer-based CodePTMs \cite{guo2020graphcodebert,wang2021codet5,wang2023codet5+} have significantly propelled the development of code intelligence. Existing CodePTMs can be mainly categorized into two classes: struct-free and struct-based. In the struct-free category, \citet{feng2020codebert} first proposes CodeBERT, a bimodal pre-trained model based on RoBERTa, which leverages masked language modeling and replaced token detection to learn NL-PL bimodal information. \citet{ahmad2021unified} follows the BART architecture and is pre-trained with denoising autoencoding on Python/Java code and NL corpus. In the struct-aware category, GraphCodeBERT \cite{guo2020graphcodebert} builds upon CodeBERT and incorporates the inherent structure of code by utilizing data flow to enhance code representation. CodeT5 \cite{wang2021codet5} builds on a unified encoder-decoder architecture based on T5 \cite{raffel2020exploring} and supports various downstream tasks and multitask learning. It introduces a novel identifier-aware pre-training task that considers the identifiers from code. UniXcoder \cite{guo2022unixcoder} adapts UniLM \cite{dong2019unified} and enhances code representation by employing AST and code comments, using mask attention matrices with prefix adapters to control the model's behavior.
\citet{xu2022survey} has found that CodePTMs with learned structural knowledge have more powerful code representation capabilities. We pursue this research line, adding structural knowledge in the fine-tuning phase.

\subsection{Neural Networks With Code Structure}
In addition to the CodePTMs discussed in Section~\ref{rela:codeptms} that attempt to incorporate structural-related pre-training tasks during the pre-training process, other studies have explored different approaches for introducing code structural knowledge.
\citet{DeepM} proposed DeepM to measure code maintainability by exploiting the lexical semantics and the structural features of text in source code.
UnixCoder\cite{guo2022unixcoder} and Hybrid-DeepCom\cite{Hybrid-DeepCom} take traversed ASTs as inputs during the pre-training phase. However, this approach is not applicable for fine-tuning, as the inputs for fine-tuning and inference should ideally align.
\textit{Great} \cite{hellendoorn2020global}, \textit{Code Transformer} \cite{zugner2021language}, and TPTrans \cite{peng2021integrating} incorporate structural knowledge into the attention computation, enabling the model to integrate program context and structure information. \textit{Great} utilizes edge type embedding in the program graph, \textit{Code Transformer} leverages node distances across the AST, and TPTrans employs relative and absolute path encoding. However, these Transformer-based methods require heavy model modifications. As a result, they cannot be readily transferred to existing pre-trained models and are not suitable for implementation during the fine-tuning phase. To address this issue, we incorporate structural information through the structure loss and multi-task fine-tuning, enabling a plug-and-play approach for our method.

\subsection{Fine-tuning Approaches for Pre-trained Models}
Existing methods for fine-tuning pre-trained models can be categorized into five types: vanilla fine-tuning, intermediate fine-tuning, multi-task fine-tuning, parameter-efficient fine-tuning, and prompt-based fine-tuning \cite{kalyan2021ammus}. For programming language processing, some researchers \cite{bogomolov2022assessing,wang2022no,ayupov2022parameter} have explored the fine-tuning of CodePTMs based on the methods above. 
\citet{bogomolov2022assessing} conducted vanilla fine-tuning under different settings and observed that performance improves as the training data becomes more relevant to the task domain. \citet{wang2022no} conducted prompt-based fine-tuning on CodePTMs and observed better results than vanilla fine-tuning. \citet{ayupov2022parameter} addressed the issue of deploying large-scale pre-trained models in integrated development environments by employing parameter-efficient fine-tuning. 
\citet{sun2023transcoder} endeavored to capture the inherent connections between different programming languages and code-related tasks to enhance the efficacy of fine-tuning.
Recently,
\citet{Y-tuning} proposed an efficient yet effective paradigm to adapt frozen large-scale PTMs to specific downstream tasks by learns dense representations.
To the best of our knowledge, there has yet to be any prior research on applying multi-task fine-tuning specifically for code pretraining models. In this study, we fine-tune CodePTMs using the multi-task fine-tuning approach.
\section{Conclusion}
In this paper, we propose SAT, a novel multi-task fine-tuning approach that enhances the structural knowledge of CodePTMs during the fine-tuning phase. We quantify the difference between the attention matrix from the Transformer block and the distance matrix extracted from the AST as the structure loss using Sinkhorn divergences. By employing multi-task learning, CodePTMs can learn code structural knowledge during the fine-tuning phase. This approach does not require modifying the model architecture, enabling SAT to possess strong portability.
To validate the effectiveness of SAT, We conducted experiments on four pre-trained models and two code generation tasks. Experimental results demonstrate that our approach enhances the representational capacity of CodePTMs and benefits tasks in low-resource scenarios.
\section{Limitations}
SAT focuses on code-to-code and code-to-text tasks but cannot be applied to text-to-code tasks. For encoder-only models, we can only leverage the structural knowledge absorption of the source sequence during fine-tuning, while we cannot absorb structural information from the target sequence due to the absence of a corresponding attention matrix. In our future work, we will explore structural enhancement methods suitable for text-to-code tasks.
\section{Ethical Considerations}
The proposed method does not exhibit potential risks. All the scientific artifacts used or created in this study are duly cited and properly licensed, and their usage aligns with their intended purpose.
\section{Acknowledgments}
\label{sec:acknowledges}
This work has been supported by the National Natural Science Foundation of China under Grant No. 62377012.

\nocite{*}
\section{Bibliographical References}\label{sec:reference}

\bibliographystyle{lrec-coling2024-natbib}
\bibliography{lrec-coling2024-example}


\end{document}